\RequirePackage{fix-cm}
\documentclass[twocolumn]{svjour3}          
\smartqed  
\usepackage{graphicx}
\usepackage{flushend}
\usepackage{hyphenat}
\usepackage{booktabs}
\usepackage{tabularx}
\usepackage{multirow}
\usepackage{subcaption}
\usepackage{float}
\usepackage{enumitem}
\usepackage{caption}
\usepackage{pifont}

\usepackage[english]{babel}   
\usepackage{hyphenat}         
\usepackage{url}
\usepackage{breakurl}
\usepackage{hyperref}

\journalname{International Journal of Information Security}
\begin{document}

\title{Enabling End-to-End APT Emulation in Industrial Environments: Design and Implementation of the SIMPLE-ICS Testbed}
\titlerunning{Design and Implementation of the SIMPLE ICS Testbed}        

\author{Yogha Restu Pramadi\and
        Theodoros Spyridopoulos\and
        Vijay Kumar
}


\institute{ Y.R. Pramadi \and T. Spyridopoulos \and V. Kumar \at 
            School of Computer Science and Informatics \\
            Cardiff University\\
            Tel.: +44-29225-10176\\
            \email{pramadiyr@cardiff.ac.uk}           
}

\date{}

\maketitle
\begin{abstract}
Research on Advanced Persistent Threats (APTs) in industrial environments requires experimental platforms that support realistic end-to-end attack emulation across converged enterprise IT, operational technology (OT), and Industrial Internet of Things (IIoT) networks. However, existing industrial cybersecurity testbeds typically focus on isolated IT or OT domains or single-stage attacks, limiting their suitability for studying multi-stage APT campaigns. This paper presents the design, implementation, and validation of SIMPLE-ICS, a virtualised industrial enterprise testbed that enables emulation of multi-stage APT campaigns across IT, OT, and IIoT environments. The testbed architecture is based on the Purdue Enterprise Reference Architecture, NIST SP 800-82, and IEC 62443 zoning principles and integrates enterprise services, industrial control protocols, and digital twin–based process simulation. A systematic methodology inspired by the V model is used to derive architectural requirements, attack scenarios, and validation criteria. An APT campaign designed to mimic the BlackEnergy campaign is emulated using MITRE ATT\&CK techniques spanning initial enterprise compromise, credential abuse, lateral movement, OT network infiltration, and process manipulation. The testbed supports the synchronised collection of network traffic, host-level logs, and operational telemetry across all segments. The testbed is validated on multi-stage attack trace observability, logging completeness across IT, OT, and IIoT domains, and repeatable execution of APT campaigns. The SIMPLE-ICS testbed provides an experimental platform for studying end-to-end APT behaviours in industrial enterprise networks and for generating multi-source datasets to support future research on campaign-level detection and correlation methods.


\end{abstract}


\section{Introduction}
\label{sec_intro}

Industrial Control Systems (ICS) and Operational Technology (OT) form the backbone of modern critical infrastructure, supporting essential services across sectors such as energy, manufacturing, water treatment, and transportation. These systems automate and optimise industrial processes to ensure operational continuity, safety, and efficiency. The ongoing digitisation and interconnectivity of ICS environments enhance productivity and flexibility but also expand the attack surface, increasing both the likelihood and potential impact of cyber incidents. Consequently, the resilience of these vital systems faces significant cybersecurity challenges.

Among the most sophisticated and persistent threats to these environments are Advanced Persistent Threats (APTs). APTs are highly coordinated systematic campaigns conducted by organised threat actors over extended periods, typically with the aim of sabotage, espionage, or theft of intellectual property. Notable ICS-focused threat actors such as Sandworm, Xenotime, and OilRig have demonstrated the capability to cause physical damage, disrupt critical operations, and compromise safety systems through carefully orchestrated multi-stage attacks. Unlike opportunistic cyberattacks, APTs are designed to evade conventional security mechanisms, maintaining stealth within a network for prolonged durations while gradually advancing toward strategic objectives. Their multi-stage nature, persistence, and exploitation of zero-day vulnerabilities make detection particularly difficult. Furthermore, the increasing convergence of Information Technology (IT) and OT, accelerated by the Industrial Internet of Things (IIoT), allows an initial compromise in the IT domain to serve as a pivot point for lateral movement into the OT environment, posing direct risks to safety, process integrity, and availability.

OT systems differ fundamentally from IT systems, prioritising availability, integrity, and operational safety over confidentiality \cite{cuorvo_securing_2024}. They often rely on legacy devices with outdated firmware \cite{hurst_chapter_2024}, lack encryption \cite{shaikh_new_2024}, and use weak or even absent authentication mechanisms \cite{nosouhi_towards_2024}. These characteristics, combined with extended system lifecycles and 24/7 operational requirements, create an environment where traditional IT security measures are insufficient or incompatible. Testing cybersecurity mechanisms or simulating complex APT campaigns on live industrial systems is impractical due to potential operational disruptions, financial losses, and physical hazards.

To overcome these constraints, realistic and reconfigurable testbeds have become indispensable tools for exploring, developing, and evaluating cybersecurity controls in ICS environments without affecting live operations \cite{ani_design_2021,fla_cybersecurity_2025}. Such testbeds provide controlled settings for vulnerability analysis, taxonomy development, and validation of novel detection mechanisms. However, a persistent challenge lies in the scarcity of representative datasets suitable for APT detection, particularly those capturing IT/OT/IIoT convergence and multi-domain correlations. Despite increasing research activity in ICS cybersecurity, publicly available datasets that support the study of multi-stage APT campaigns in converged IT-OT environments remain limited. Most existing datasets focus on isolated attack techniques, single network domains, or short time windows, and often lack the temporal fidelity, contextual detail, and coordinated enterprise, network, and operational telemetry required to model stealthy, long-lived APT behaviour across heterogeneous industrial systems \cite{conti_survey_2021,saxena_advanced_2023}. As a result, researchers frequently rely on custom-built testbeds to generate experimental data tailored to specific threat models and detection objectives. High-fidelity testbeds capable of producing synchronised, cross-layer telemetry across the full APT lifecycle are therefore critical for enabling rigorous analysis, reproducible experimentation, and systematic evaluation of detection and defence mechanisms in converged industrial environments.

Despite this, few existing platforms enable the safe and reproducible study of APT campaigns spanning the full enterprise architecture from IT to OT and IIoT. Many current testbeds focus on isolated domains, lack the architectural complexity of modern industrial networks, or provide limited instrumentation for cross-domain event correlation. These constraints hinder the development and evaluation of detection mechanisms capable of identifying the subtle, coordinated activities that characterise advanced threat actors as they move laterally through segmented environments. Furthermore, the proprietary nature and extensive customisation requirements of most existing testbeds reduce their reproducibility and limit their adaptability across different industrial sectors.

This gap motivates the development of the proposed \textbf{Simulated Industrial Multitier Platform for Laboratory Emulation of Industrial Control Systems (SIMPLE ICS)} testbed, which enables systematic emulation of multi-stage APT campaigns and comprehensive data collection in converged industrial environments. The acronym SIMPLE reflects an aspirational design goal: while constructing a realistic industrial cybersecurity testbed is inherently complex, this work aims to provide a reproducible, modifiable, and sector-agnostic platform that can be adapted to diverse industrial contexts with minimal effort. The SIMPLE ICS testbed is designed as a model-based experimental environment to support the study of APT behaviours and the generation of high-fidelity datasets for the evaluation of detection and defence mechanisms in converged industrial networks.


The main contributions of this paper are summarised as follows:
\begin{itemize}
    \item \textbf{Integrated IT--OT--IIoT testbed architecture}: 
    We present a sector-agnostic industrial cybersecurity testbed architecture that realistically models modern enterprise environments by integrating IT, OT, and IIoT domains with layered network segmentation, industrial protocols, and enterprise services aligned with the Purdue model.

    \item \textbf{End-to-end APT emulation framework}: 
    We introduce a reproducible methodology for emulating multi-stage APT campaigns that traverse enterprise IT, OT, and IIoT environments, grounded in documented threat actor behaviours (e.g., Sandworm, Xenotime, OilRig) and MITRE ATT\&CK for ICS.

    \item \textbf{Comprehensive, synchronised data collection}: 
    We establish a framework for synchronised multimodal data collection across network, host, and process layers, forming a foundation for advanced APT detection research aemulatednd providing a template for data set generation.

    \item \textbf{Systematic testbed development methodology}: 
    We adopt the V-Model design methodology to ensure traceability from research requirements to implementation and validation, supporting reproducibility and structured evaluation of the proposed testbed.
\end{itemize}


The remainder of this paper is structured as follows: Section~\ref{sec_related_works} reviews related works and highlights existing gaps. Section~\ref{sec_methodology} describes the adaptation of the V-Model methodology to develop and validate the testbed. Section~\ref{sec_apt_ien} discusses the progression of the APT attack in industrial enterprise networks and introduces a generalised APT scenario. Section~\ref{sec_testbed} details the design and implementation of the SIMPLE ICS testbed. Section~\ref{sec_evaluation} evaluates its performance. Section~\ref{sec_limitations} outlines limitations and future directions, and Section~\ref{sec_conclusions} concludes the paper.


\section{Related Works}
\label{sec_related_works}

The research landscape surrounding industrial network security against APT attacks has progressed significantly in recent years, largely driven by increasing threats and regulatory pressures. However, a crucial observation emerges: a dominant architectural trend -particularly in established testbeds- overwhelmingly focuses on replicating the OT or IIoT network environment. While valuable, this approach suffers a significant limitation: it largely neglects the significance of the IT infrastructure in facilitating and enabling APT attacks within industrial settings. This section will review existing research, highlighting the shortcomings of siloed ICS testbeds and arguing for a more holistic, integrated approach.

APTs in industrial environments are characterised by multi-stage campaigns that traverse layered network architectures while maintaining stealth during lateral movement. Traditional signature-based or single-stage anomaly detectors are insufficient for identifying such threats, as they typically focus on isolated artefacts (e.g., anomalous flows or known signatures) rather than correlated long-term behaviours. \cite{choi_comparison_2019} emphasised the need for detection methods that leverage multi-source correlation, temporal context, and cyber threat intelligence to capture campaign-level activities rather than discrete attack events . A recent literature review of APT detection in ICS highlight challenges such as heterogeneous data sources, IT/OT interconnectivity, and the absence of longitudinal datasets that represent the entire lifecycle of APT \cite{abu_talib_apt_2022,al-kadhimi_systematic_2023}.

To address these challenges, researchers have employed multistage models, such as the Cyber Kill Chain and MITRE ATT\&CK, to correlate alerts across different phases of an attack. Techniques include probabilistic graphical models \cite{shen_data-driven_2020,cuong_using_2023}, sequence mining \cite{bhattarai_prov2vec_2023,abu_talib_apt_2022}, and graph-based or temporal correlation frameworks \cite{al-aamri_machine_2023}. These approaches enhance context-awareness by linking seemingly disparate events into coherent attack traces. However, their effectiveness depends heavily on access to realistic, multi-stage, and well-labelled datasets resources that remain scarce due to industrial confidentiality and data sensitivity. Consequently, research increasingly relies on complex custom built testbeds to generate representative datasets \cite{myneni_unraveled_2023}.

Several ICS cybersecurity testbed projects have developed physical, virtual, or hybrid testbeds to support the generation of data sets and the evaluation of detection techniques. Survey works have catalogued such efforts \cite{conti_survey_2021}, encompassing process-control testbeds, smart grid simulators, water treatment facilities, hardware-in-the-loop (HIL), and digital twin platforms. However, the survey discussed that most of the testbeds focus on single attacks (e.g., denial-of-service, command injection) rather than the multi-stage, cross-domain nature of APT campaigns. Even when multi-phase attacks are included, they often remain confined within the OT domain.

In 2021, Al-Hawawreh and Sitnikova \cite{al-hawawreh_developing_2021} presented Brown-IIoTbed, a hybrid testbed (using raspberry pis and arduinos) focused on IIoT brownfield systems. The testbed blends legacy system with modern IIoT and its architecture follows the three-tier architecture. The testbed shows high fidelity as an IIoT testbed but it does not follow any cybersecurity standard to provide baseline security. And the attack scenario implemented are non-APT cyber attacks and assumes that the attacker has gain a foothold within the network. Architecturally, the testbed is not designed to represent a full industrial enterprise network with IT networks and common IT services. This makes it not suitable to recreate certain attacks such as lateral movement that exploit \textit{Active Directory} services that are common in APT attacks.

Ravikumar et al. \cite{ravikumar_cps_2021} presented a hybrid testbed specifically for Wide Area Monitoring, Protection and Control (WAMPAC), a network commonly used in electrical power systems to control grid operations. It used a real-time digital simulator (RTDS) hybridised with real industry-grade remote terminal units (RTU) and substation automation systems (SAS). The use of an RTDS makes the testbed achieve a high degree of fidelity in process simulation and data while simulating a very large industrial network, but at the same time these testbeds are not easy to recreate because of the high cost of these simulators, and special training is needed to use them. The cyber attacks implemented on the testbed cover a specific scenario of attacks targeting power grids, and not focused on advanced multi-staged attacks.

Mubarak et al. \cite{mubarak_industrial_2021}, built a portable physical testbed designed to generate data sets for machine learning. The testbed focused on the acquisition of real-time OT data and attacks that impact OT operations, making its architecture only limited to the OT network. This also limits the attacks that can be implemented, resulting in a dataset that only has representation of ICS hacking attacks.

Karch et al. \cite{karch_crosstest_2022}, presented CrossTest, a physical cybersecurity-focused testbed with miniaturised power systems and a miniaturised manufacturing process. The testbed implemented a cross-layer ICS testbed that integrates network, host, and control-layer monitoring to align with APT stages. Although this improved the richness of the data set, the testbed remained limited to OT operations, lacking the IT-to-OT breach vector.

Kumar and Thing \cite{kumar_raptor_2023}, taking inspiration from the Brown-IIoTbed, presented a similar testbed to detect APT attacks in the IIoT environment but with the addition of using real PLC hardware as the controller. They presented three APT scenarios attacking IIoT with scenarios mimicking real-world APT attacks on industrial environment. Similarly, Ghiasvand et al. \cite{ghiasvand_cicapt-iiot_2024} also building upon Al-Hawawreh’s IIoT testbed, adopted the design for a provenance-based APT attack detection approach. The study implemented more than 20 APT attack techniques. These testbed also inherent the short comings of the Brown-IIoTbed.

Simola et al. \cite{simola_developing_2023}, developed a cyberphysical testbed to assess the effectiveness of SOCin critical infrastructures. The project was designed with the focus on mobile technology such as 4G and WIFI in industrial settings and only focussing on the OT network. The study briefly discusses the array of ICS centric cyber attacks, but does not discuss how it is implemented.

Mikkelsplass and Jorgensen \cite{mikkelsplass_cyber_2023} introduced a physical testbed using a scaled-down factory model controlled by a real PLC. Similarly to the previous work, the study investigates how effective SOCs are in detecting cyber attacks in complex OT settings. The network represents an ICS network with an IIoT gateway without any representation of the IT network and the other layer of the IIoT architecture which limits the scope of attacks.

Lo et al. \cite{lo_digital_2024} presented a virtual testbed that utilises digital twin technology (DT). The  study used a factory simulator in conjunction with an open-source PLC. The aim of the study is to generate time series operational data for Deep Learning based detection. The study shows that DT can provide enough fidelity to study the impacts of cyber attacks on the process layer; nevertheless, as with the other studies in this section, the scope is limited on the OT network.

In summary, there remains a lack of reproducible high-fidelity testbeds that model the full Purdue enterprise architecture encompassing the IT, OT, and IIoT layers. This limitation restricts adversary emulation and hinders the study of end-to-end APT campaigns that traverse enterprise IT to the operational domain and ultimately impact physical processes. The objectives outlined in Section~\ref{sec_intro} directly address these gaps.

\section{Methodology}
\label{sec_methodology}
The development of a cybersecurity testbed for industrial networks is a complex undertaking that requires a methodical and rigorous approach. The credibility of any research conducted on such a testbed is intrinsically linked to the credibility of the testbed itself, which in turn depends on the process used for its design and construction \cite{green_ics_2020}. This section establishes the foundational principles for the testbed development lifecycle, justifying the selection of methodology used as the guiding framework.

\begin{figure}[h]
\includegraphics[width=0.5\textwidth]{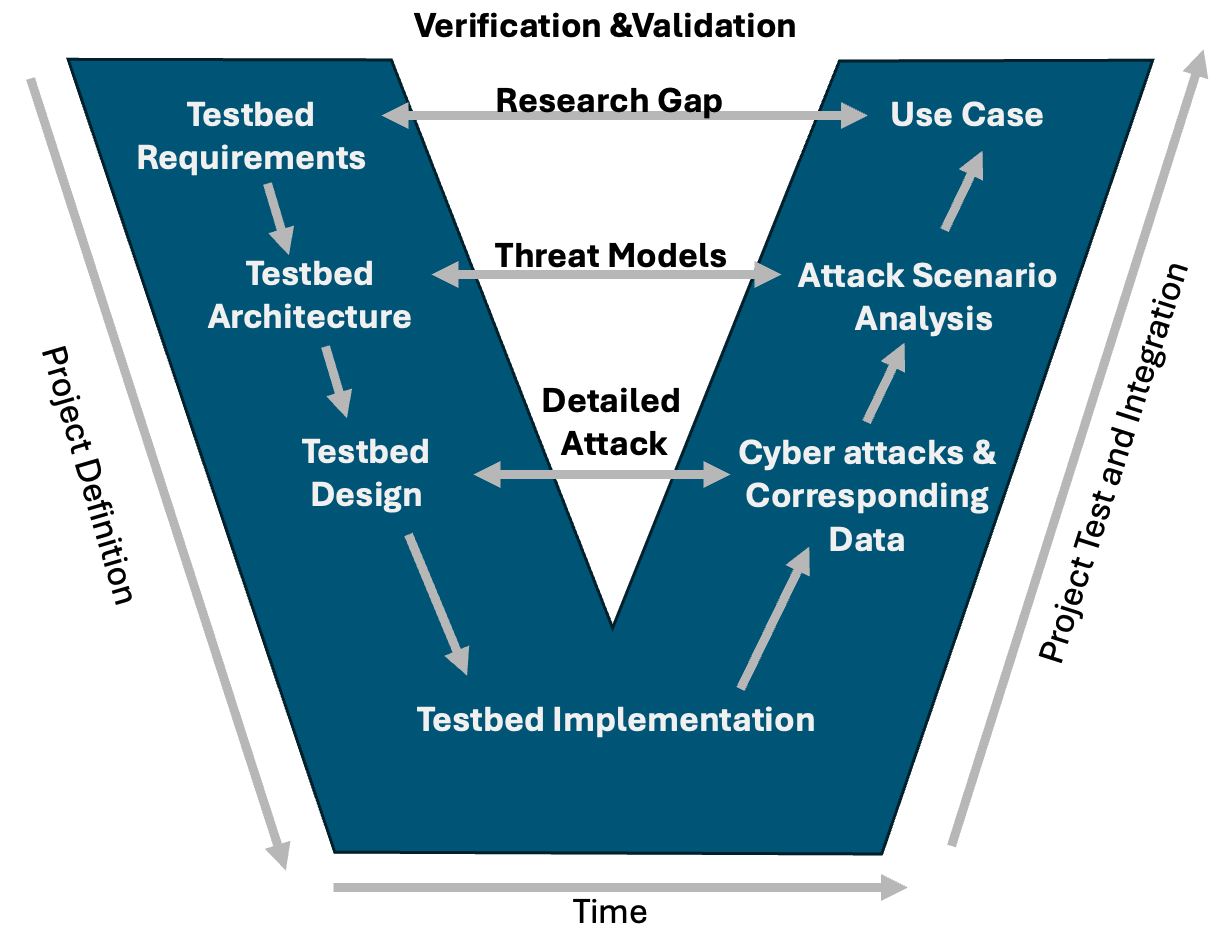}
\caption{Proposed V-Model Design Methodology for cybersecurity testbed development}
\label{v_model}
\end{figure}

This research adopts the V-Model Design Methodology \cite{pressman_software_2010}, commonly used in system engineering processes and software engineering \cite{dogan_v-model_2021,durmus_enhanced_2018}, to guide the development of a testbed for the detection and evaluation of APTs in industrial enterprise  networks. This model is characterised by a sequential development process in which each phase of specification and decomposition is mirrored by a corresponding phase of integration and validation. Figure \ref{v_model} shows the model adapted for the study. 

The left side of the 'V', representing the decomposition and specification, mirrors the process commonly used in the construction of ICS cybersecurity testbeds \cite{ani_design_2021}.  It begins with the high-level \textbf{Requirements}, which are analogous to the overarching research questions. Then these are decomposed into a system \textbf{Architecture}, which represents the high-level experimental design. Finally, the \textbf{Design} phase specifies the concrete components and configurations, defining the experimental apparatus and procedures.

The right side of the 'V', representing integration and testing, mirrors the process of experimental verification and hypotheses testing (validation). First, the \textbf{Testbed Use Case} corresponds to the defined testbed use cases that are derived from the \textbf{research gap} that the research is trying to address . Next, the \textbf{Attack Scenario Analysis} is based on \textbf{the threat model} or the threat scenario of real-world cyberattack scenarios, and the testbed architecture must be in line with the intended scenario it recreates. The last is the \textbf{ cyberattack testing and data collection validation} , where each important component of the designed testbed is tested for its functionality, simulating the attack, testing data collection capability, and performance to ensure completeness and quality of the data. 

Finally, the two sides merge into the \textbf{Implementation} phase. The testbed is implemented, validated and verified by performing the APT attack simulation as specified by the generalised APT scenario and evaluating the generated data. Validating that the fully integrated system, the testbed, can successfully run the experiments needed to answer the initial research questions. This parallel structure establishes clear and logical verification and traceability from the highest-level research objective down to the lowest-level component test and back up to the final validation.

The usage of the methodology in this study is as follows:
\begin{itemize}
    \item \textbf{The Requirement - Testbed Use Case Phase:} The methodology begins with the identification of the testbed requirements based on a gap analysis of existing testbeds review of industrial APT incidents in the literature as discussed in \ref{sec_related_works}. 
    \item \textbf{The Architecture - APT Scenario Simulation Phase:} The requirements and cybersecurity standards (e.g., NIST SP 800-82, IEC 62443) are then used to determine the architecture and validated by realistic use cases formulated that reflect known APT campaigns. Three attack scenarios are a generalisation of prominent APT attacks in the industrial environment. These scenarios span the entire APT lifecycle, from initial access to impact, and are mapped to MITRE ATT\&CK techniques to guide the functionality of the testbed and the log requirements. The architecture can be found in Subsection \ref{sec_testbed_architecture}, and the attack scenarios are detailed in Subsection \ref{sec_apt_scenario}.
    \item \textbf{The Design - APT Attacks \& Corresponding Data Phase:}The design specifies component interconnections, logging configurations, and attack simulation mechanisms are discussed in Section \ref{sec_testbed_design}. Both IT and OT-specific telemetry sources are instrumented, including PCAPs, OT protocol logs, Windows event logs, and endpoint activity. The telemetry are then mapped to their corresponding attacks and discussed in Section \ref{sec_apt_scenario_mapping}. 
    \item \textbf{The Implementation Phase:}The final implementation is implemented in a virtualized environment using open source platforms such as OpenPLC, Factory I/O, Node-RED, VirtualBox, and Proxmox (detailed in Section \ref{sec_testbed_implementation}. Infrastructure-as-Code tools (e.g. Ansible and Terraform) are used to enable reproducible deployment. The testbed supports the full execution of each selected APT scenario, providing comprehensive logs and facilitating live monitoring during red-team simulations. Implementation is validated through the phased execution of APT campaigns, confirming that each attack technique produces observable evidence in the data layer (Section \ref{sec_evaluation}).
\end{itemize}
This systematic methodology ensures that the SIMPLE ICS testbed is developed with rigour, traceability, and reproducibility.

\section{Developing the APT Threat Model}
\label{sec_apt_ien}

\begin{figure}[h]
    \centering
    \includegraphics[width=8cm]{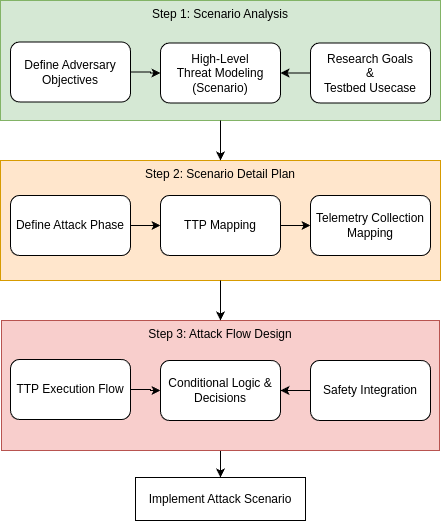}
    \label{fig:subfigure_scenario1}
    \caption{The Approach to Design a Detailed APT Scenario for Cybersecurity Testbed }
    \label{fig_apt_methodology}
\end{figure}

This section aims to discuss the design of a detailed APT attack threat modelling in the context of industrial cyber security studies. This is a crucial step in the study as it serves as the basis of the testbed design and drives the validation and verification steps of the methodology described in Section \ref{sec_methodology}. To help achieve this, the study devised an approach aimed at creating detailed APT attack scenarios specifically for cybersecurity testbeds. The approach is illustrated in Figure \ref{fig_apt_methodology}.

The approach and its usage in this study is described in detail below.

\textbf{Step 1: Scenario Analysis}. The threat model scenario is developed by taking into account two things, the research goals and the adversary objectives that the research is trying to investigate. The research goals mandate the main objective of the study, for example, "generating a dataset of APT attacks with lateral movement from IT to OT". This goal defines the type of adversary with a threat model that fits the study best. From this example, the study focusses on detecting APT attacks with \textit{lateral movements}, then choosing to mimic the \textit{Sandworm} APT group for a scenario is a logical choice. From this, a high-level scenario threat model is developed, as can be observed in Subsection \ref{sec_apt_scenario} .  

\textbf{Step 2: Scenario Detail Plan}. In this step, the scenario from the first step is broken down into detailed attack phases and mapped to MITRE ATT\&CK TTPs. Once the detailed TTP map is finished, the telemetry that responds to the attack is mapped to help the process of collecting data for each attack. This step is used to guide Subsection \ref{sec_apt_scenario_mapping}.

\textbf{Step 3: Attack Flow Design}. This step is crucial in relation to the implementation of the attack. Each attack in the scenario has an execution flow to follow in order for an attack to succeed. And the result of each attack has an impact that may concern any safety issues to which the testbed must adhere. This step creates conditional logic of the attack and the decisions that to ensure realism, fidelity, and the safety of the execution of the performed APT attack scenario. Subsection \ref{sec_apt_implementation} describes this step in detail.

This section is structured as follows. In the first section, the use case of the testbed is derived from the research gap; The second subsection describes the APT threat model in an industrial context; And the last subsection breaks down the APT scenario into a detailed attack mapped to its.

\subsection{A Generalised APT Scenario Targeting ICS}
\label{sec_apt_scenario}

To help verify the second step in the v-model methodology that we addopt, \ref{sec_methodology}

The testbed requirement requires it to support the simulation of APT attacks in industrial environments.  This means that the APT scenarios that it needs to simulate are APT campaigns that specifically target industrial networks (e.g., ICS or IIoT) infrastructure as their primary objective, not just a minor component of broader IT attacks against organisations with industrial networks. This subsection discusses studies and reports on prominent APT campaign analysis and presents a distillation of these APT campaigns into a generalised APT scenario.

A generalised APT scenario provides a repeatable, explainable threat model to guide the testbed design: it defines the adversary goals, the stages or phases of the campaign, the detailed scenario of chain of attacks, and the stages where detection and mitigation controls must operate.

To effectively describe an APT scenario, it is crucial to break it down into sequential phases that illustrate the progression of an attack. APTs targeting industrial enterprise networks follow a structured attack path, typically encompassing reconnaissance, initial compromise, persistence, privilege escalation, lateral movement, command and control (C2), and final objectives such as operational disruption or data exfiltration. Frameworks like the ICS Cyber Kill Chain \cite{assante_industrial_2021}, the MITRE Attak for ICS \cite{alexander_mitre_2020} and the Mandiant Attack Lifecycle \cite{maynard_decomposition_2020} provide a structured approach to studying these threats. 

\begin{figure}[h]
    \centering
    \includegraphics[width=8cm]{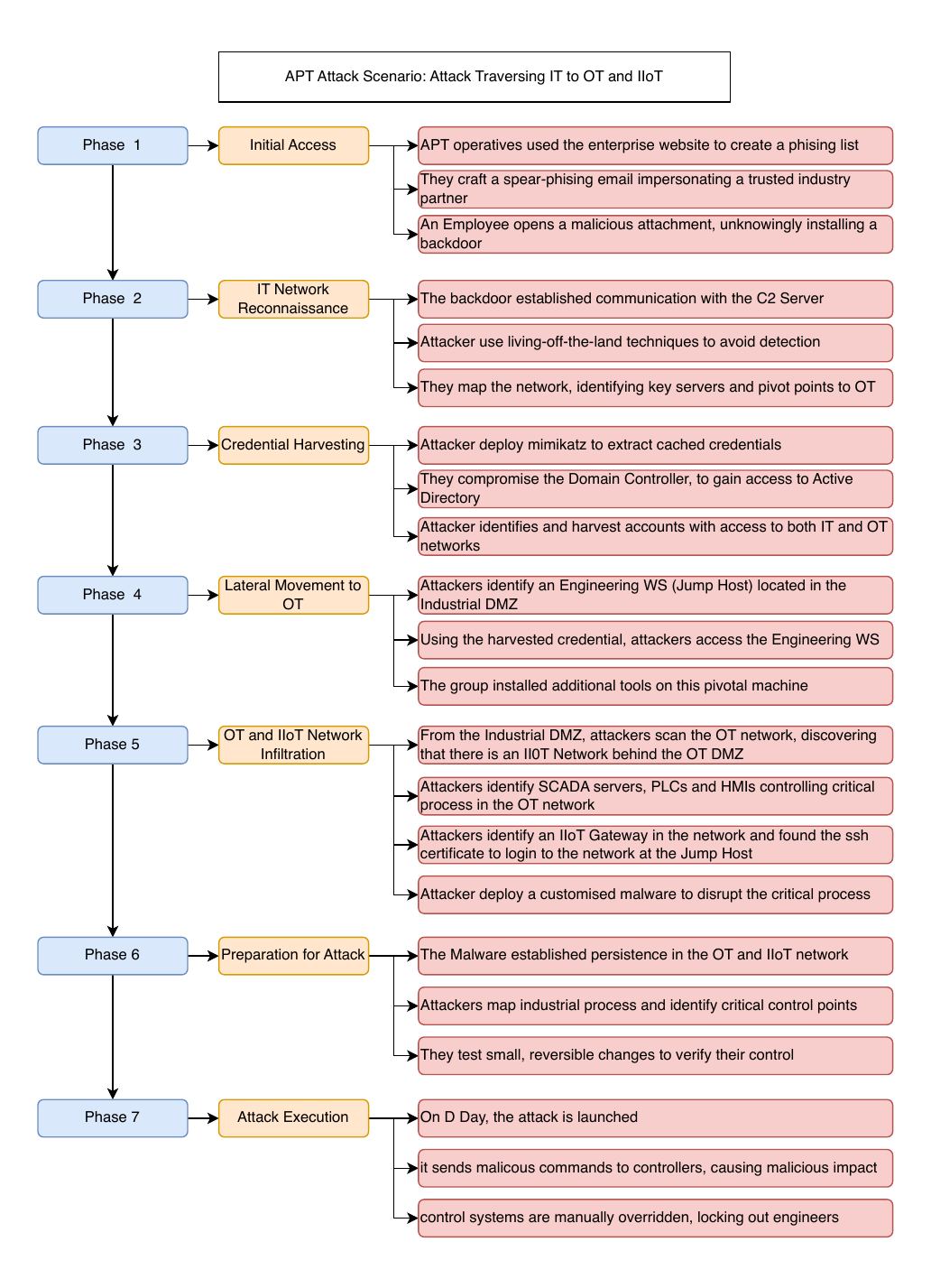}
    \label{fig:subfigure_scenario1}
    \caption{A Generalisation of APT attack scenarios in industrial environments. }
    \label{fig_apt_scenario}
\end{figure}

This study presents a generalised APT scenario (depicted in Figure \ref{fig_apt_scenario}) that represents prominent APT campaigns. The scenarios are developed by analysing studies such as \cite{alladi_industrial_2020,maynard_decomposition_2020,kumar_apt_2022,khan_threat_2016,stojanovic_apt_2020}. The scenario presents an APT campaign with the mission of compromising a critical process in IT and OT. Inspired by the 2016 and 2022 attacks on the Ukraine's power grid by Sandworm (APT 28). The scenario starts with a spear-phishing campaign for it's initial access phase, then the attack traverses the IT network to the OT network by using harvested credential to access the jump host in the OT DMZ. In the OT DMZ the attack identifies industrial equipment in the OT and IIoT network, and conducted an attack causing disruption to a critical industrial process. A detailed process of this scenario is shown in Figure \ref{fig_apt_scenario}.

The APT scenario is developed to simulate an advanced attack that targets modern OT networks that include IIoT networks. Although there is no single publicly documented APT campaign that demonstrated such an attack, this scenario is synthesised from studies such as \cite{kumar_raptor_2023} and \cite{ghiasvand_cicapt-iiot_2024}. 

\subsection{From Scenario to TTPs}
\label{sec_apt_scenario_mapping}
The APT attack scenarios from the previous section provides a high abstraction of series of correlated attacks. To make it actionable, complex attack chains in a scenario are translated into discrete tactics, techniques, and procedures (TTP) \cite{yang_tactics_2023}. MITRE ATT\&CK \cite{alexander_mitre_2020} provides a common vocabulary and taxonomy to describe adversary behaviours to facilitate a clearer description of the attack involved. This subsection maps the generalised attack campaign from the previous step of the study to TTPs, providing a clear blueprint for the design phase of the testbed to follow.

The process of mapping the generalised scenarios into the MITRE ATT\&CK TTP framework is as follows:
\begin{enumerate}
    \item Understand the Scope of the Attack: The MITRE ATT\&CK differentiate between IT enterprise, ICS, Mobile, and Cloud matrices. For IT/OT scenarios, the enterprise and ICS matrices are used. The scenarios are reviewed and the attack phase is defined within the IT and OT scope.
    \item Deconstruct the Scenario into Atomic Attacks: The phases of attacks are then broken down into steps of attacks (tactics), and the the steps are detailed into the smallest action or events possible (techniques and subtechniques). The produced discrete attack must follow the chronological flow and how one action enables the next.
    \item Analyse Each Action Against ATT\&CK Tactics: The next step is to identify the corresponding technique and sub-technique in the matrices for each discrete action and codify them accordingly. 
    \item Refine and Organise the Mapping: The last step is to organise the identified TTPs according to the attack phases described in the scenario. The mapped scenarios must align with the progression of the attack.
\end{enumerate} 

Table \ref{table_mapped_ttps} presents the mapping of the first scenario. The first column describes the phase of the attack, the second column describes the tactics, and the granular attack techniques and subtechniques described in the last column. Note that in \textit{phase 1} to \textit{phase 5} the tactics and techniques use the enterprise matrices, and from \textit{phase 6 }onwards it uses the ICS matrices. The first five phases detail the atomic steps that the adversaries took in compromising the IT enterprise network and make their way to the edge of the OT network, and the next three phases detail the adversaries' activity in the OT network up to the execution of the final objective. 

\begin{table*}[!t]
\caption{Detailed APT Scenario}
\label{table_mapped_ttps}
\centering
\scriptsize
\renewcommand{\arraystretch}{1.15}
\setlength{\extrarowheight}{0pt}
\begin{tabularx}{\textwidth}{p{3cm} X p{3.5cm} p{3.5cm}}
\toprule
Phase & TTP Mapping & Asset Target & Telemetry \\
\midrule

Phase 1: Initial Access &
Initial Access (TA0001) \newline
-- Spearphishing Attachment (T1566.001) \newline
& IT user workstation, email gateway &
NIDS: Malware Transfer, HIDS/EDR: process creation, firewall logs\\[2pt]
\addlinespace[2pt]
\midrule

Phase 2: IT Network Reconnaissance &
Command \& Control (TA0011) \newline
-- Standard Application Layer Protocol (T1071.001) \newline
Discovery (TA0007) \newline
-- System Information Discovery (T1082) \newline
-- Network Service Discovery (T1046) \newline
Defense Evasion (TA0005) \newline
-- Impair Defenses (T1562)
& Active Directory servers, IT user workstation &
NIDS, Network Flow: C2, HIDS/EDR: sysmon event logs, SIEM correlation \\[3pt]

\addlinespace[3pt]
\midrule

Phase 3: Credential Harvesting &
Credential Access (TA0006) \newline
-- OS Credential Dumping (T1003) \newline
---- LSASS Memory (T1003.001) \newline
Discovery (TA0007) \newline
-- Account Discovery (T1087) \newline
---- Domain Account (T1087.002) \newline
-- Network Share Discovery (T1135)
& Active Directory servers, IT user workstation &
 HIDS/EDR: Windows security logs (4624, 4625, 4672), LSASS memory access logs. \\[3pt]

\addlinespace[3pt]
\midrule

Phase 4: Lateral Movement to OT &
Lateral Movement (TA0008) \newline
-- Remote Services (T1021) \newline
---- SMB Admin Shares (T1021.002) \newline
Execution (TA0002) \newline
-- Valid Accounts (T1078) \newline
& Jump servers, engineering workstations, file shares, OT gateway &
NIDS: network scan,Authentication logs, SMB and RDP connection attempts, file transfer logs, Sysmon (Event 3, 10), network flow \\[3pt]

\addlinespace[3pt]
\midrule

Phase 5: OT and IIoT Network Infiltration &
Discovery (TA0008 ICS) \newline
-- Device Discovery (T0808) (ICS) \newline
-- Process Discovery (T0813) (ICS) \newline
Execution (TA0007 ICS) \newline
-- Command-Line Interface (T0807) (ICS) \newline
Persistence (TA0006 ICS) \newline
-- Boot or Logon Autostart Execution (T1547) \newline
& PLCs, HMIs, IIoT gateways, engineering stations &
PLC communication logs, Modbus/TCP network captures, HMI command logs, OPC-UA telemetry, host startup logs \\[3pt]

\addlinespace[3pt]
\midrule

Phase 6: Preparation for Attack &
Collection (TA0100) \newline
-- Monitor Process State (T0801) \newline
Inhibit Response Function (TA0011 ICS) \newline
-- Change Credential (T0892 ICS) \newline
& SCADA servers, PLCs, process controllers, operator HMIs &
NIDS:ICS anomaly detection logs, Operational logs, Process historian data \\[3pt]

\addlinespace[3pt]
\midrule

Phase 7: Attack Execution &
Impair Process Control (TA0009 ICS) \newline
-- Modify Parameter (T0836 ICS) \newline
Impact (TA0010 ICS) \newline
-- Manipulation of View (T0832 ICS) \newline
-- Manipulation of Control (T0831 ICS) 
& Field controllers, actuators, sensors, safety systems &
PLC change logs, process control commands, historian integrity data, ICS alarms, safety PLC telemetry \\

\bottomrule
\end{tabularx}
\end{table*}


\section{Testbed Architecture, Design and Implementation}
\label{sec_testbed}
This section discusses the architecture, design, and implementation of the proposed testbed. 
\subsection{Architecture and Design of the Testbed}
\label{sec_testbed_architecture}
\begin{figure}[h]
\begin{minipage}{\columnwidth}
\includegraphics[width=1\linewidth]{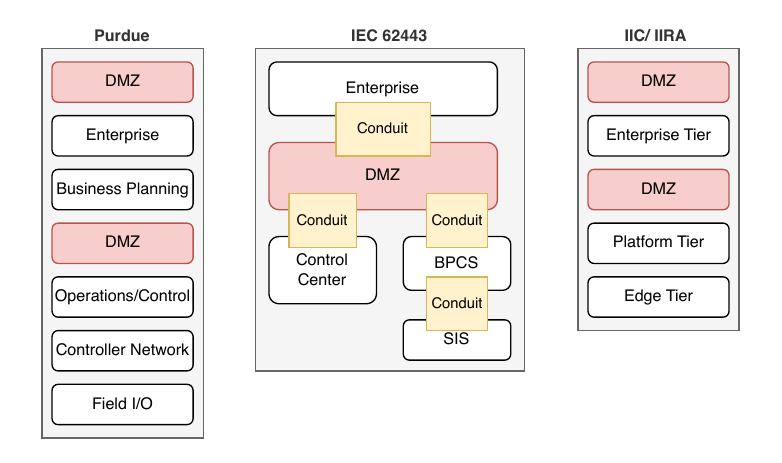}
\end{minipage}
\caption{Industrial Network Architectures According to \cite{stouffer_guide_2023} 
\label{fig_architectures}}
\end{figure}

In Section \ref{sec_related_works} we have discussed the gaps in ICS testbeds for APT research and developed a set of requirements to guide the architecture development. To fulfil the requirement of representing a modern industrial enterprise network that includes IT network, OT network, and IIoT network, we used the Purdue Enterprise Reference Architecture (PERA) in conjunction with the Industrial Internet Reference Architecture (IIRA) as a guide. To provide baseline security, the IEC 64223 recommendation is followed to implement security zones and no direct IT to OT communication is followed. Figure \ref{fig_architectures}.

PERA provides a robust foundation due to the inherent layered approach, focusses on industrial processes, and established applicability across diverse industrial sectors \cite{li_formalization_1994,garton_purdue_2019}. It is built around a six-level hierarchical model, each representing increasing abstraction from the physical plant floor to the enterprise level. \textbf{Levels 0-1 (Physical Process)} represent the actual equipment, sensor, actuators, and controllers on the factory floor. \textbf{Level 2-3 (Control and Operations)} is where the supervisory control and data acquisition systems reside, managing the physical process underneath them. \textbf{Levels 4-5 (Business/Enterprise)} represent the IT environment in which the business applications and the corporate network reside. 

For the IIoT portion of the design, IIRA provides a framework that allows interoperability between various technologies in the industrial domain \cite{al-hawawreh_developing_2021}. IIoT is characterised by a wide array of devices, protocols, and platforms from different digital industrial generations and vendors. The IIRA enables the modelling of the complex ecosystem by utilising a layered abstraction architecture that are tailored to serve specific needs of industrial application. The framework is also data-centric with emphasis on data management and analytics, and it encourages the use of open standards (e.g., MQTT, OPC UA, AMQP) for communication and data exchange between layers. This allows for cybersecurity experimentation focused on data integrity and process-based anomaly detection. The testbed design follows this framework by adopting the three-tier model.
\begin{itemize}
    \item \textbf{IIoT Edge Tier:} The edge tier represents the core OT environment devices and systems that handle the physical process and the systems that translate and communicate data and control with the tier above it.
    \item \textbf{IIoT Platform Tier:} This tier acts as a data aggregator, data storage, and data and control relay between the edge layer and the enterprise layer.
    \item \textbf{Enterprise Tier:} The enterprise level is the layer that consumes the operational data and presents it to the business stakeholders to aid in monitoring and decision making. This tier can also provide direct control to certain physical devices enabling smart processing control and remote maintenance.
\end{itemize}

With the use of PERA and IIRA as a reference for the testbed architecture, a realistic representation of a contemporary industrial enterprise network can be achieved.  NIST 800-82 and IEC 64223 also contribute to the design by providing baseline security measures and security data monitoring points to capture multimodal and high-fidelity security data.
\subsubsection{Industrial Process Design}

Physical OT devices and physical industrial processes on testbeds are desirable for their high fidelity, but often create a barrier for the researcher to implement a testbed for cyber security research. For certain research such as anomaly detection based on operational data, using high-fidelity data from real physical devices and physical processes is the best approach. This study will focus more on how APT attacks move and operate within an industrial enterprise network and create a defence mechanism as early as possible. This does not necessitate the need for a high-fidelity source of physical process data. However, abstract models or simplified representations of OT may lead to oversimplification. Oversimplification can cause the missing of critical data due to the lack of realistic interactions and interdependencies.

The integration of Digital Twin (DT) technology into cybersecurity testbeds has emerged as a transformative approach to modelling and simulating complex cyberattack scenarios in virtual environments. By creating high-fidelity virtual replicas of physical systems, DTs enable organisations to test and analyse cybersecurity threats without risking real-world infrastructure. 

Digital Twins offer several advantages when integrated into cybersecurity testbeds:
\begin{itemize}
    \item High-Fidelity Virtual Models: DTs create detailed virtual replicas of physical systems, allowing for accurate simulations of cyberattack scenarios. For example, in smart cities, DTs can model critical urban infrastructure to simulate complex attacks \cite{gulyamov_using_2024}. Similarly, in industrial systems, DTs can replicate the behaviour of industrial control systems (ICS) to evaluate novel cyber-defence strategies \cite{cuorvo_securing_2024}.
    \item Cost-Effective and Safe Testing: Conducting cybersecurity tests in virtual environments is safer and more economical than testing in real-world settings. For example, in hydroelectric power plants, DTs allow for comprehensive security tests without risking operational disruptions \cite{erkek_enhancing_2024}.
    \item Advanced Threat Simulation: DTs enable the simulation of various attack scenarios, including sophisticated Advanced Persistent Threats (APTs). In smart manufacturing, DTs can simulate the impact of various threat scenarios to train deep learning models for threat detection \cite{lo_digital_2024}. 
\end{itemize}
With these benefits, DT technology is chosen as the approach that the testbed will use to represent the industrial process.
\subsection{Testbed Design}
\label{sec_testbed_design}
This section describes the design of the testbed based on the architectures discussed in the previous section. The design also must be able to accommodate the simulation of the APT scenarios defined in Section \ref{sec_apt_scenario}. This is done by identifying the required network segments, services, and devices that the testbed must have.
To facilitate a full attack scenario simulation with scenarios that contain a pivot of the attack from IT to OT, the test bed is designed with six segments. The distinct network segments are as follows:
\begin{enumerate}
    \item \textbf{Internet Simulator and Cyber Attack Platform (External Network):} a network segment that is not part of the simulated industrial network, but plays a vital role in providing internet service simulation (e.g. simulated DNS, web, and routing). This segment will also be the origin of the attacks that are conducted by the simulated APT.
    \item \textbf{IT DMZ (Demiliterized Zone):}  A buffer zone between the IT LAN and external networks (e.g. the Internet). It hosts services that require external access (e.g. Active Directory, email, webservices, and other general business applications) while protecting internal IT assets. In accordance with the IIRA architecture, this segment also hosts the enterprise tier of the IIoT network.
    \item \textbf{IT LAN:} Represents the corporate enterprise network, including typical office workstations. This segment is the most exposed to external threats and often serves as the initial point of compromise for APTs.
    \item \textbf{Industrial/OT DMZ:} This critical buffer zone is positioned between the IT and OT networks and is protected by a firewall on both connections. It hosts devices that require interaction with both IT and OT environments, such as data historians, data log servers, and various gateways, ensuring controlled data flow and limiting direct IT-to-OT communication. The OT DMZ serves as a single, carefully controlled point of entry to the OT network, often accessed via a jump host.
    \item \textbf{OT Network:} This segment houses the core industrial control systems, including SCADA systems, PLCs, HMIs, and other intelligent devices responsible for monitoring and managing physical processes. This network prioritises availability and safety, using specialised industrial communication protocols. Micro-segmentation within the OT network further isolates devices into functional zones, minimising impact of breach.
    \item \textbf{IIoT Network:} This infrastructure integrates IIoT devices that connect to the OT network and the enterprise networks, facilitating operational monitoring and efficiency. These devices often lack robust security protocols, making them potential weak links for lateral movement between IT and OT. The IIoT network design within the testbed reflects its role as a potential target for attackers.
\end{enumerate}

To better mimic real-world conditions, the testbed is also designed with best practices and cyber security guidelines, such as NIST SP 800-82 \cite{stouffer_guide_2023} and IEC62443 \cite{noauthor_isaiec_nodate}. These standards emphasise robust segementation and secure conduits to control traffic flow and minimise attack surfaces. The architectural design of the testbed incorporates multiple layers of defence.
\begin{itemize}
    \item \textbf{Dual Firewalls:} Strategic placement of firewalls at the edge of the IT network and the OT network, with the OT DMZ between them, enforces access policies and controls traffic flow. These firewalls are configured to perform packet filtering and enforce micro-segmentation rules within the OT and IIoT environment.
    \item \textbf{Network Segmentation and VLANs:} The entire simulated enterprise network is divided into distinct IT and OT domains, with further segmentation into Virtual Local Networks (VLANs) within each domain. This allows for tailored security measures and different access policies, reducing exposure to unsophisticated attack vectors.
    \item \textbf{Access Control:} A strict access controls using technology commonly used such as Access Directory.
    \item \textbf{Jump Hosts:} A common security best practice is to have a jump host or a bastion host. A jump host serves as the point of entry for external users to access devices on the OT network from the IT network. This minimises entry points and ensures that only authenticated and authorised personnel can cross over into the OT domain.
\end{itemize}

\begin{figure*}[h]
\includegraphics[width=15 cm]{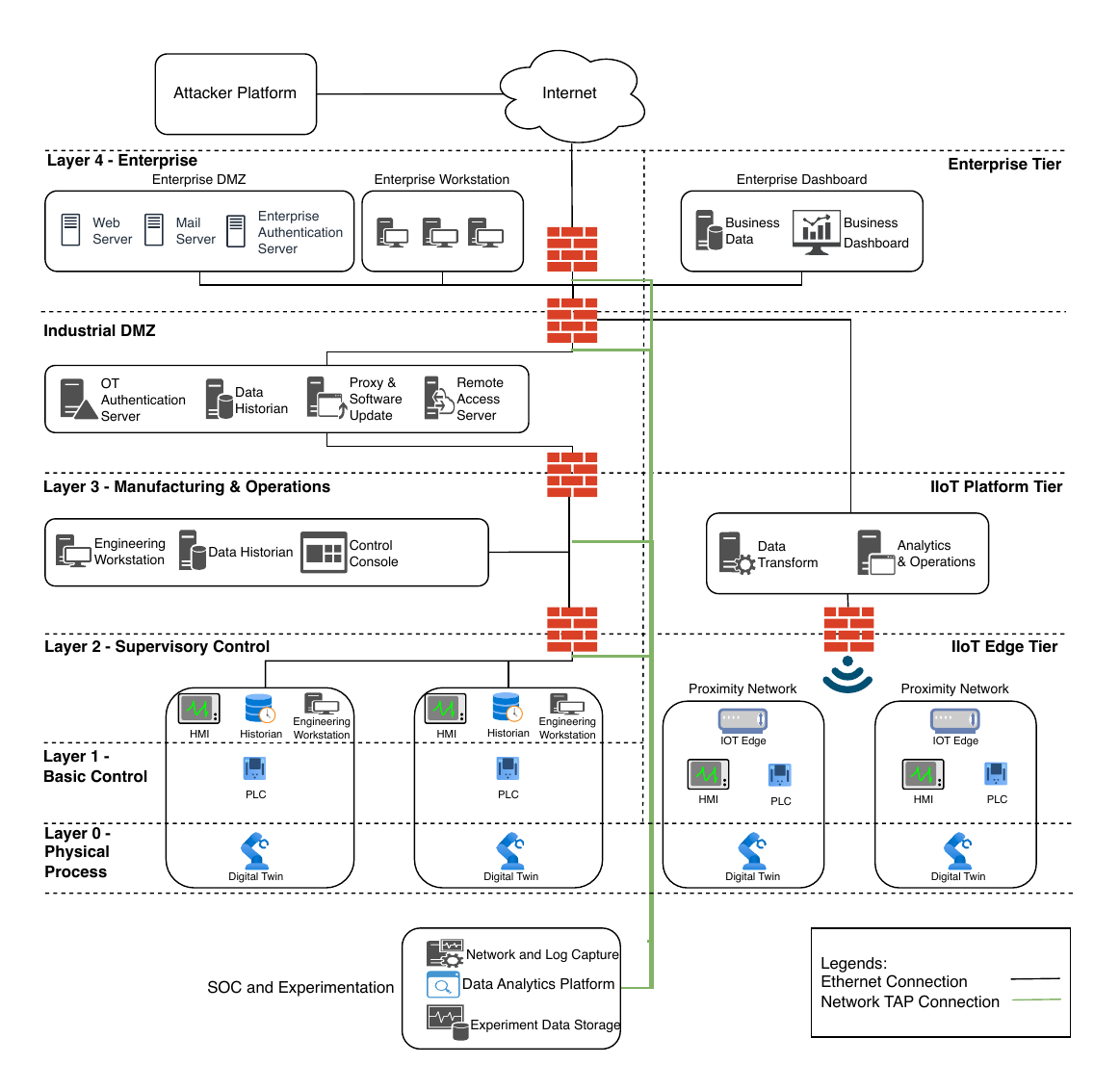}
\caption{The Proposed Testbed Design, Guided by PERA and IIRA \label{fig_testbed_design}}
\end{figure*}

Figure \ref{fig_testbed_design} illustrates the design of the testbed with its network segments, network devices, virtual machines, virtualised OT devices, and digital twins. 

\subsection{Testbed Implementation}
\label{sec_testbed_implementation}
To implement the testbed, we follow the practical implementation guide presented by \cite{green_ics_2020}. The study presents a building block of a cybersecurity testbed for industrial networks, the building blocks are the management layer, the user layer, the infrastructure bridge, and the experimental layer. Providing these layers is a big task if it is done from scratch. Cyber range platforms such as \cite{noauthor_ludus_nodate,lieskovan_building_2021,noauthor_cyber_nodate} provide the building blocks and tools required to accelerate the building process of a testbed. For this study, we chose the DIATeam cyber range \cite{noauthor_cyber_nodate}.

\begin{figure}[h]
    \includegraphics[width=0.5\textwidth]{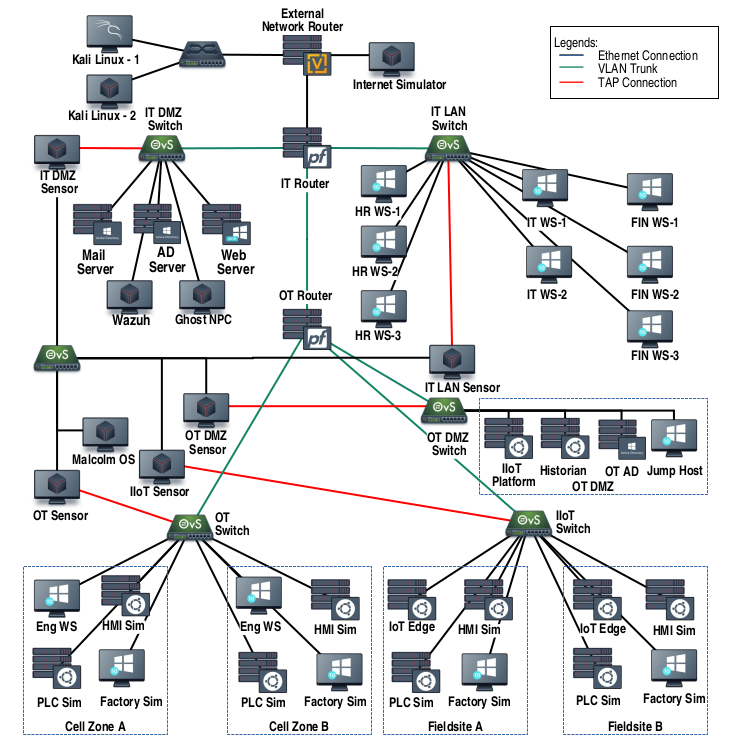}
    \caption{Testbed Implementation on DIATeam
    \label{fig_testbed_implementation}}
\end{figure}

The test bed is implemented with mostly open source software, with the proprietary software used being Factory I/O, Microsoft Windows 10 Education and Microsoft Windows Server. Figure \ref{fig_testbed_implementation} depicts the overall testbed topology implemented in the DIATeam Cyberrange. A brief description of the testbed implementation is discussed as follows.

Implementing the testbed starts with building the emulated network using open source firewalls, routers, and switches. The \textbf{main network infrastructure} is built with a \textit{VyOS} router, two \textit{pfsense} firewalls, and five \textit{OpenVswitch} virtual switches. The \textit{VyOS} router serves as a router to route outbound packets from the internal network to the attacker platform and the internet simulator. The first \textit{pfsense} router acts as the perimeter firewall with four network interfaces, an interface for the IT DMZ, an interface for the IT LAN, and an interface for the backbone that connects to the OT network. The IT LAN interface on the firewall is configured to host intranet VLANS. The second \textit{pfsense} firewall acts as an OT firewall that regulates the traffic entering and leaving the OT network. The OT firewall has four network interface, an interface to the enterprise backbone, and interface to the OT DMZ, an interface to the OT network, and an interface to the IIoT network. The firewalls are configured to provide basic services such as DHCP, NAT (Network Address Translation), and baseline segmentation security. Each firewall interface is connected to an \textit{OpenVSwitch} virtual switch configured to forward packets to the first port for packet monitoring, and for the switches that serves VLAN segmentation, the switches interface are configured accordingly.

The implementation of the \textbf{External Network Segment} is built with two main components, the internet simulator and the attack platform. The internet simulator uses \textit{TOPGEN} and \textit{GreyBox}, we used the readily available OVA and converted it to a QCOW2 (QEMU Copy-On-Write) format. The VM is configured with a bridge interface that will be the destination of traffic that is intended to be directed to the internet, the Greybox emulates the internet backbone, while TOPGEN provides functional equivalents to popular internet services such as HTTP, DNS, and email. The attack platform uses \textit{Kali Linux} and \textit{Parrot OS} with Caldera along with other common hacking tools configured to simulate APT attacks.

For the \textbf{IT Enterprise DMZ Segment}, two Windows Server are used. The first Windows Server VM serves as the main Active Directory configured as the domain controller (DC) of a domain (simpleics.local). The AD is configured with vulnerabilities and misconfiguration that enable lateral movement, taking inspiration from the \textit{Games of Active Directory} (GOAD) project. The implemented vulnerabilities are: \textit{Kerbroasting}, \textit{ASREPRoasting}, Unsecured SMB shares, and \textit{Golden Ticket} persistence. The second server is installed with an email service and IIS to provide a platform for a vulnerable web service and a platform to deliver phishing emails.

The \textbf{IT LAN Segment} implements multiple VMs using Windows 10 configured with the Wazuh agent and the Ghosts NPC human behaviour agent. The VMs are authenticated using the domain controller, and the human behaviour agent runs a looped human behaviour scenario and APT victim scenarios. The human behaviour simulation is described further in section \ref{sec_normal_behaviour}.

\unskip

The \textbf{OT DMZ Segment} hosts an OT subdomain domain controller (\textit{ot.simpleics.local}) and two IIoT platform servers. The DC uses \textit{Windows Server 2019}, configured as a trusted child sub domain controller. The IIoT platform is built with Ubuntu Server running the MING (\textit{MQTT}, \textit{InfluxDB}, \textit{Node-RED}, and \textit{Grafana}) stack. The MING stack serves as a broker with four main functions: firstly it receives published messages from the IIoT gateway and relay them to the enterprise platform, secondly it receives control messages from the enterprise platform and forward them to the IIoT gateways, thirdly it stores operational telemetry data  . 

The \textbf{OT Segment} implements two Windows 10 VMs and two Ubuntu Desktop 24.04. The first Windows VM plays a role of an Engineering Workstation with OpenPLC Editor installed, the second VM serves as a factory simulator using Factory IO. The Ubuntu Desktop VMs are dedicated to simulate industrial control devices, the first VM servers as a PLC using OpenPLC, and the second VM simulates an HMI using FUXA. 

The \textbf{IIoT Segment} is implemented similarly to the OT segment, the difference is that this segment uses the MING stack at the IIoT edge gateway instead of using Modbus. The MQTT service acts as a client that publish operational messages to the IIoT platform in the OT DMZ  and subscribes to topics that control the factory process. This enables remote control of the process from the IIoT Dashboard. 

The last segment to implement is the \textbf{Experiment/SOC Segment}, this segment is implemented with a segregated network using an OpenVSwitch virtual switch as the main network hub of this segment. To gather log data we deployed five Hedge Hog Linux VM as a network sensor, one for each network segment in the testbeds network (not including the external segment). The network sensor has two interface, one acts as the network tap interface and the other is for connecting to the main SOC platform. The main SOC platform uses Malcolm OS to process the gathered logs and packets from the network sensors, it acts as a central analysis platform of the experimental segment of the testbed. Table \ref{tab_tools_components} presents a list of applications, tools, and operating systems chosen for the testbed.

The SimpleICS implementation, configuration files, and scripts used in this study are publicly available in our repository \cite{pramadi_pramadiyrsimpleics_nodate}.

\begin{table*}[!t]
\caption{Tools and Components.}
\label{tab_tools_components}
\centering
\small
\renewcommand{\arraystretch}{1.2}
\begin{tabularx}{\textwidth}{p{2.3cm} p{3cm} p{3.2cm} X}
\toprule
\textbf{Segment} & \textbf{Virtual Device} & \textbf{Tools} & \textbf{Role} \\
\midrule

\multirow{3}{*}{External Network} 
 & VyOS & -- & External network routing \\
 & Parrot OS and Kali VM & MITRE Caldera & Adversary simulator \\
 & Internet simulator VM & TopGen, GreyBox & Internet simulator \\
\midrule

\multirow{4}{*}{IT DMZ} 
 & PFSense & -- & IT enterprise firewall and routing \\
 & OpenVSwitch & -- & Virtual switch and network tap \\
 & Windows Server 2016 & Active Directory & Authentication server, web server, and mail server \\
 & Ubuntu Server & MQTT, InfluxDB, NodeRed, Grafana & IIoT enterprise platform \\
\midrule

\multirow{2}{*}{IT LAN} 
 & OpenVSwitch & -- & Virtual switch and network tap \\
 & Windows 10 & Ghost NPC & Emulates human behaviour and serves as an attack pivot point \\
\midrule

\multirow{3}{*}{OT DMZ} 
 & PFSense & -- & OT network firewall and routing \\
 & Windows Server 2016 & Active Directory & Authentication server, web server, and mail server \\
 & Ubuntu Server 24.04 & MQTT, InfluxDB, NodeRed, Grafana & IIoT platform \\
\midrule

\multirow{2}{*}{OT Network} 
 & Windows 10 & FactoryIO & Factory process simulator \\
 & Ubuntu Desktop 24.04 & OpenPLC, FUXA & Open-source PLC and HMI \\
\midrule

\multirow{3}{*}{IIoT Network} 
 & Ubuntu Server 24.04 & MQTT, NodeRed, InfluxDB & IIoT gateway \\
 & Windows 10 & FactoryIO & Factory process simulator \\
 & Ubuntu Desktop 24.04 & OpenPLC, FUXA & Open-source PLC and HMI \\
\midrule

\multirow{3}{*}{SOC Network} 
 & OpenVSwitch & -- & Virtual switch and network tap \\
 & Malcolm OS & Open-source network analysis toolset* & Network traffic analysis tool suite \\
 & HedgeHog Linux & Open-source network analysis toolset* & Network PCAP log \\
\bottomrule
\end{tabularx}

\noindent{\footnotesize{* A detailed list of the open-source toolset can be seen in \cite{noauthor_malcolm_nodate}.}}
\end{table*}

\subsubsection{Industrial Process Simulation Implementation}
As discussed in the previous section, this study opted to use DT technology to simulate factory processes mainly to facilitate deployment and reproducibility while maintaining high fidelity to the simulation of physical processes for cybersecurity research purposes. Factory IO provides such a tool at a relatively affordable price. To aid in reproducibility, the study used the readily available scenes in Factory IO. For controlling and monitoring the process, the study opted to use free open source solutions, namely OpenPLC for the PLC simulator and FUXA for the HMI/SCADA simulator. The logic implemented for the scene is implemented using the IEC 611131-3 Ladder Diagram (LD) Standard. The implementation of the industrial process for the OT and IIoT segments is as follows:
\begin{enumerate}
    \item \textbf{The OT Segment Industrial Process Implementation:} The industrial process in this segment mimics a metal lid production line using the "Production Line" scene from Factory IO. The process involves loading metal slabs from a conveyor belt to a CNC machine (computer numerical control) using a robotic arm, lid production control, production progress monitoring, and transport to the end of the production line with conveyor belts. In the simulation, there are two production lines that work in parallel, with the end of the conveyor merging into one exit. The ladder diagram takes input from the diffuse sensors for the conveyor and the progress status of the production process. Then it controls the scene by sending a control signal to the conveyor belts and the CNC machine. The HMI implements the monitoring of the process, it displays every status of the conveyor, counts the produced lids from each station, and implements three manual controls (start, stop, and reset). The SCADA presents the same information and control with the addition of presenting and storing the process information over time (daily) to the historian database.
    \item \textbf{The IIoT Segment Industrial Process Implementation:} This segment uses the "Sorting by Weight" scene, which simulates a boxed package sorting system that sort by the weight of the box. The process starts by sensing the weight of the box using a conveyor scale and determines the weight of the box by measuring the output voltage of the sensor (0 - 10 volts). The box will then be redirected to three different conveyors depending on the weight of the box. The ladder diagram of this scene takes input from diffuse sensors of the conveyors to sense the position of the box; then it senses the voltage of the scale sensor, calculates the box's weight, and sends signals to the pop-up wheel sorter depending on redirecting the box to the intended conveyor. Similarly to the OT segment, the HMI and SCADA for the IIoT process displays every input and output of the process, presents additional controls to manage the process, and stores process data to the historian.   
\end{enumerate}

The simulated industrial process uses the Modbus TCP/IP protocol, as this protocol is considered the de facto protocol for communication within ICS \cite{jhan_enhancing_2023}. And for IIoT communication, the MQTT protocol is used for its ubiquitous adoption in IIoT due to its lightweight and efficient nature that requires minimal resources to run \cite{al-hawawreh_developing_2021}. In this setup, the Factory IO acts as a Modbus TCP/IP server, and OpenPLC, FUXA, and Node-Red act as the clients. Taking note of this configuration, to address the modbus outputs and inputs, it is necessary to follow the OpenPLC addressing space for a \textit{slave} configuration: discrete outputs (\%QX100.0-\%QX199.7) and discrete inputs (\% IX100.0-\%IX199.7). Similarly in FUXA, the input needs to be offset by 800 bits and the output offset by 800 bits.

\begin{figure}[htbp]
\centering
\begin{subfigure}{0.45\textwidth}
    \centering
    \includegraphics[width=\linewidth]{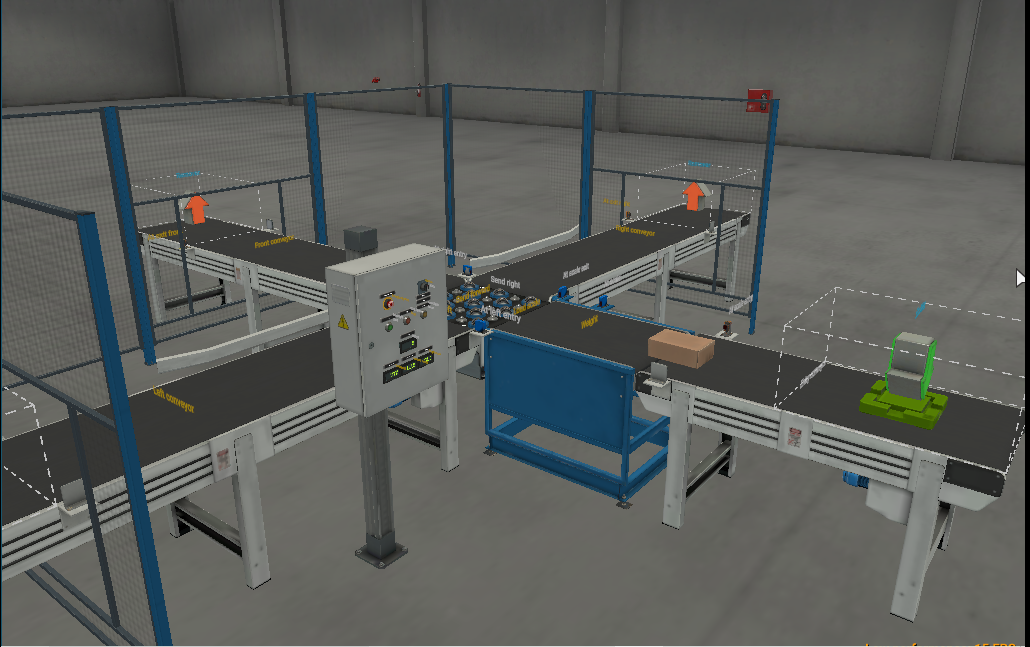}
    \caption{Factory Simulation with Factory I/O}
    \label{fig:factoryio}
\end{subfigure}
\hfill

\begin{subfigure}{0.45\textwidth}
    \centering
    \includegraphics[width=\linewidth]{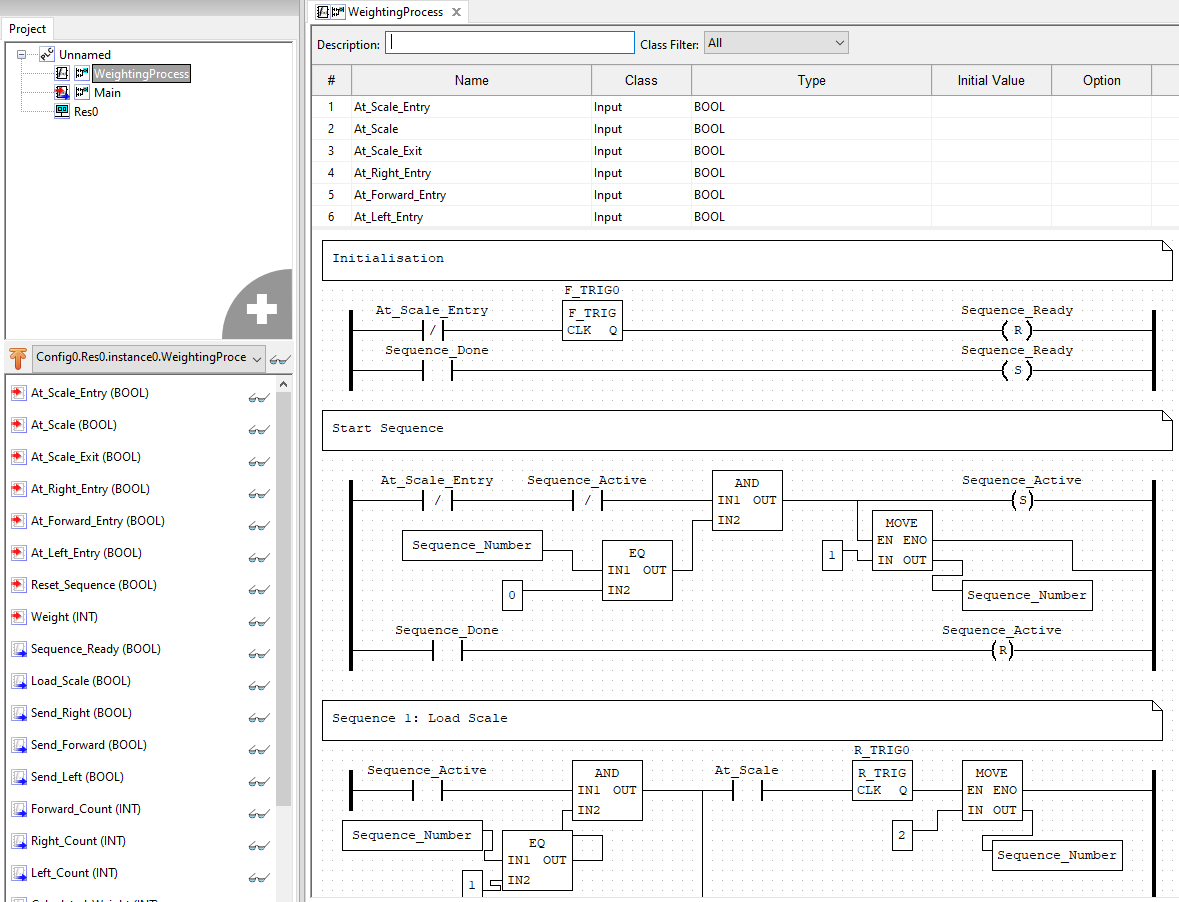}
    \caption{Ladder Diagram}
    \label{fig:ladder_diagram}
\end{subfigure}
\hfill

\caption{The developed OT environment}
\label{fig_screenshot}
\end{figure}

Figure \ref{fig:factoryio} shows the "Sorting by Weight" scene used in the IIoT segment, and Figure \ref{fig:ladder_diagram} shows an example of the ladder diagram for the "Sorting by Weight" scene developed using the OpenPLC Editor.

\subsection{Simulating User and Operational Behaviour}
\label{sec_normal_behaviour}
In a complex network such as an industrial network, normal behaviour data can be divided into three categories: normal human activity behaviour, normal operational behaviour, and background noise. This requirement is vital for effective anomaly detection, where any significant deviation from the established baseline can signal suspicious behaviour or potential compromise. The subsection covers human behaviour simulation tools for the testbed, focussing on the Ghosts NPC tool.

Ghosts NPC is a tool that automates the simulation of human behaviour using agents (NPCs) and a centralised API to manage and monitor the behaviour. They use game theory and concepts to animate the agents by giving the agents a persona that mimics complex human behaviour. The latest iteration of the tool supports the use of generative AI to animate agents and will have a non-deterministic outcome.

In this study, to simulate normal human / user behaviour, the "knowledge horizon" of normal human activities is followed \cite{updyke_ghosts_2018}.  This means that to construct a normal user behaviour in a testbed we need to know what the role of the user is, identify what the user will realistically know and behave, the timing of the actions of the user (when they will act and how long does it take), and the interaction between users within an organisation. IT and OT users will have similarities in performing IT related tasks, and they will behave differently relative to their job roles.

In addition to normal behaviour, the role of a user in an APT attack scenario is also identified. This can be used to simulate users with low security awareness clicking a phishing link or unknowingly activating a malware \cite{veksler_simulations_2018}. As the study focusses on APT, the scenario in this study does not include any insider threat attacks, but it can be done if necessary in the future, as the tool used permits such use cases. 

The study implemented five roles, namely normal IT user, victim IT user, engineering department user, normal OT user, and victim OT user. An interactive action between the user, which involves sending emails between NPCs, is also defined in the script. The details of the roles, activities, actions, and rationale for the activities can be seen in Table \ref{tab_simulated_behaviour}. 

\begin{table*}[h]
\caption{Simulated User Behaviour.}
\label{tab_simulated_behaviour}
\centering
\small
\renewcommand{\arraystretch}{1.12}
\setlength{\extrarowheight}{0pt}
\begin{tabularx}{\textwidth}{p{3.6cm} p{3.8cm} p{3.8cm} X }
\toprule
\textbf{Role} & \textbf{Activity} & \textbf{Actions \& Frequency} & \textbf{Justification} \\
\midrule

Normal IT User & Web browsing & Accesses company intranet and online documentation (5 times/day). & Mimics common office tasks and provides baseline network traffic. \\[4pt]

 & Document creation \& editing & Creates and saves documents in shared drives; edits existing files (3 times/day). & Represents typical productivity workflows and file I/O. \\[4pt]

 & Email communication & Sends and receives emails (10 emails/day). & Generates routine email traffic for baseline behaviour. \\

\midrule

Victim IT User & Web browsing & Same as Normal IT User, but clicks a phishing link when triggered. & Acts as a precursor to compromise. \\[4pt]

 & Email communication & Same as Normal IT User, but opens malicious attachments when present. & Simulates susceptibility to social engineering. \\

\midrule

Engineering Department (Jump Host Access) & System login \& maintenance & Logs into jump host via RDP for maintenance (2 times/day). & Reflects periodic maintenance tasks and establishes legitimate access patterns. \\

\midrule

Normal OT User (Engineer) & PLC monitoring & Uses SCADA/HMI to monitor PLC status and view sensor data (continuous during working hours). & Represents core industrial control system operation. \\

\midrule

Victim OT User (Engineer) & PLC monitoring & Same as Normal OT User but executes a compromised binary after compromise. & Introduces malicious code into OT environment. \\

\midrule

Interactive Action: Email & Victim IT User $\rightarrow$ OT User & Victim IT user accidentally forwards a potentially malicious attachment to an OT user. & Simulates accidental propagation of malware/malicious content across trust boundaries. \\

\bottomrule
\end{tabularx}
\end{table*}

\subsection{Implementing APT Scenarios}
\label{sec_apt_implementation}
This section describes the practical implementation of the APT scenario that this study proposes, leveraging open source offensive security tools and the Caldera cybersecurity framework. The goal is to emulate each phase of the attack lifecycle in an IT-OT convergence environment, allowing empirical evaluation of detection and correlation mechanisms.

To emulate the multi-phase scenario, we employ the MITRE Caldera adversary emulation platform \cite{noauthor_caldera_nodate}. Caldera's modular architecture enables the orchestration of ATT\&CK-aligned attack campaigns via agents and abilities, supporting repeatable, auditable experiments. Although Caldera does not implement all techniques in the MITRE ATT\&CK framework (especially initial access techniques to avoid being miss-used), it provides substantial coverage for enterprise and ICS TTPs by installing the official OT protocol plugins \cite{noauthor_caldera_nodate}. For techniques that are not covered by Caldera's abilities, the study implements manual attacks.

The methodology in implementing attack scenarios with Caldera in this study is as follows:
\begin{enumerate}
    \item \textbf{Planning}: This phase involves planning the simulated attack by analysing the scenarios' TTPs against the testbed topology, determining the attack path and the chain of attacks of the simulated apt. We identify the VMs that will be the victims, the vulnerability that will be exploited on the system, and the tools that will be used for the attack. This chain of TTPs will be implemented as a series of "ability".
    \item \textbf{Ability Selection and Creation}: an ability in Caldera is an individual atomic technique that corresponds to a specific attack technique. Each ability is given a TTP id number that reflects the corresponding MITRE ATT\&CK TTP.  Caldera comes with a set of default \textit{abilities}. However,as stated above, there are cases where the ability that needs to be implemented is not in the library. In such cases, we need to define a new ability; we can create them by scripting a manual command to execute an external tool.
    \item \textbf{Adversary Creation}: an adversary in Caldera is a simulated threat actor that is created by selecting a series of abilities. The abilities is organised according to the phase of attacks documented in the scenario. 
    \item \textbf{Attack Implementation}: After creating the adversary, the next phase is the attack implementation phase. In a complex scenario, full automation of attacks is difficult to achieve. This complexity arises from the interdependence of one ability on the results of a preceding ability. The steps in implementing attacks include:
    \begin{enumerate}
        \item \textbf{Agent Deployment:} an agent is a lightweight implant that executes commands on behalf of the adversary. It serves as the remote agent to simulate APT action accross the attack. Agents need to be deployed in each victim in the scenario. Deployment typically simulates Inital Access (TA0001) by installing the agent on a compromised host.
        \item \textbf{Operation Execution:} An Operation in Caldera is an orchestrated, multi-step adversary emulation campaign. It chains together individual abilities, executed by agents on compromised hosts.
    \end{enumerate}    
\end{enumerate}

Table \ref{tab_caldera_abilities} presents the adversary configuration for scenario 1. The scenario is successfully implemented with Caldera and it only requires manual intervention for the spearphishing each process. The Caldera modbus plugin is used for the final phase of the attack.

The SimpleICS-APT dataset, including PCAPs, OT process logs, and event annotations, is available at the following repository: \cite{pramadi_simple_2025}.

\begin{table*}[!t]
\caption{APT Emulation Implementation}
\label{tab_caldera_abilities}
\centering
\small
\renewcommand{\arraystretch}{1.12}
\setlength{\extrarowheight}{0pt}
\begin{tabularx}{\textwidth}{p{3.0cm} p{5.0cm} X}
\toprule
\textbf{Attack Phase} & \textbf{ATT\&CK Tactics \& Techniques} & \textbf{Caldera Abilities \& Plugins} \\
\midrule

Initial Access &
TA0001 (T1566.001, T1133) &
Manual spear-phishing with a malicious attachment or link; initial script drop to install the \texttt{sandcat} agent enabling remote HTTP(S) C2. \\[6pt]

\midrule

Reconnaissance \& C2 &
TA0011, TA0007, TA0005 &
Run discovery and environment-collection abilities such as \texttt{whoami}, \texttt{systeminfo}, \texttt{net view}, \texttt{arp}, \texttt{netstat}. \\[6pt]

\midrule

Credential Harvesting &
TA0006 (T1003) &
Execute built-in \texttt{mimikatz} abilities to dump LSASS credentials; harvested credentials are stored as Caldera \texttt{facts} for subsequent use. \\[6pt]

\midrule

Lateral Movement &
TA0008 (T1021.*, T1078) &
Use \texttt{psexec}, \texttt{smbexec} and \texttt{stockpile} abilities to pivot laterally across IT hosts to the OT-Jumphost using valid credentials on remote-services. \\[6pt]

\midrule

OT execution Preparation &
TA0007, TA0002 &
Issue shell commands on the OT-Jumphost, enumerate interfaces, probe OT subnets, and identify modbus targets. \\[6pt]

\midrule

ICS Actions &
TA0010, TA0009, T0831 &
Use the Caldera-OT plugin to modify Modbus-TCP writes to impact PLC interactions. \\

\bottomrule
\end{tabularx}
\end{table*}

\begin{figure}[htbp]
\centering
\begin{subfigure}{0.47\textwidth}
    \centering
    \includegraphics[width=\linewidth]{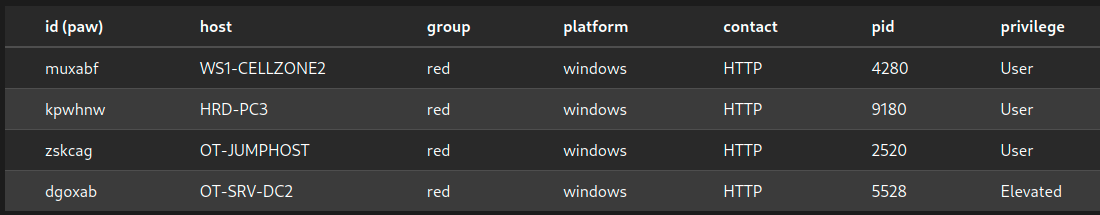}
    \caption{Compromised Machines}
    \label{fig:sub_compromised_vm}
\end{subfigure}

\begin{subfigure}{0.47\textwidth}
    \centering
    \includegraphics[width=\linewidth]{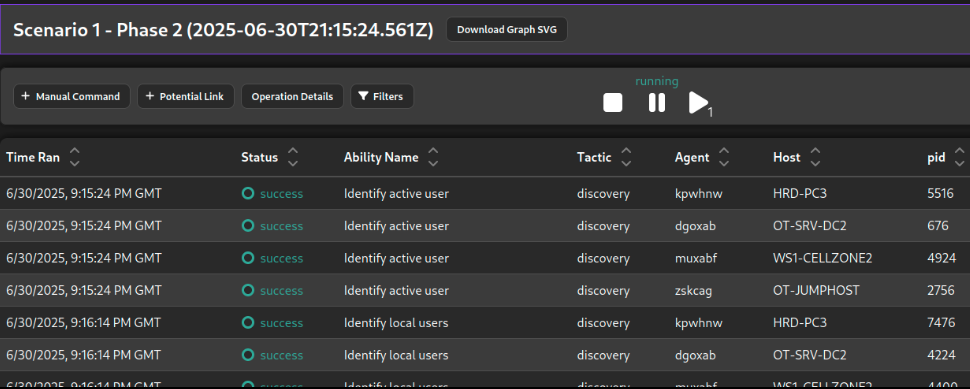}
    \caption{Automation of Phase 2 (Reconnaissance)}
    \label{fig:sub_attack_auto}
\end{subfigure}

\begin{subfigure}{0.47\textwidth}
    \centering
    \includegraphics[width=\linewidth]{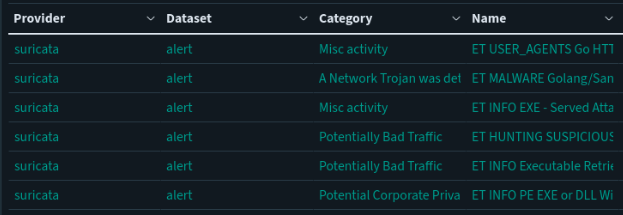}
    \caption{Detected Attacks on the Malcolm Dashboard}
    \label{fig:sub_malcolm_detect}
\end{subfigure}

\caption{Advanced Attack Simulation with Caldera: (a) Compromised Machines, (b) Automation of Scenario 1 – Phase 2 (IT Reconnaissance), (c) Detected Attacks on the Malcolm Dashboard.}
\label{fig_apt_automation}
\end{figure}


\section{Validation and Evaluation}
\label{sec_evaluation}
\subsection{Functional and Performance Evaluation}
The performance of the testbed was evaluated across three critical communication pathways: intra-IT communication, intra-OT communication, and cross-segment communication. Performance measurements were collected over a 1-day period under varying operational conditions to establish baseline characteristics and identify potential bottlenecks.

\begin{table}[htbp]
\caption{SIMPLE-ICS Performance Over One Hour}
\label{tab_net_performance}
\centering
\small
\renewcommand{\arraystretch}{1.1}
\begin{tabularx}{\columnwidth}{l *{3}{>{\centering\arraybackslash}X}}
\toprule
\textbf{Metrics} & \textbf{Intra-IT} & \textbf{Intra-OT} & \textbf{Cross-Segment} \\
\midrule
Avg. Network Latency (ms)   & 0.974  & 1.29  & 3.605 \\
Avg. Bandwidth (KB)         & 9463   & 9102  & 9012 \\
Avg. Packet Loss (Normal)   & 0      & 0     & 0 \\
Avg. Packet Loss (Stress)   & 0      & 0     & 0 \\
\bottomrule
\end{tabularx}
\end{table}

Table \ref{tab_net_performance} shows the performance. The network latency, average bandwidth, and packet loss measurements are performed over an hour to reveal the performance profiles between the different network segments. The network latency and the average bandwidth of Intra-IT communications consistently showed low latency and excellent network throughput performance with a mean response time of 0.974 ms and an average bandwidth of 9463 Kbps. Intra OT communication exhibits slightly higher latency at 1.29 ms with 9102 kbps throughput. The cross-segment network latency from the IT network to the OT network is measured as 3.695 ms and a throughput of 9012 Kbps. This shows that the network performance is well within acceptable bounds for ICS communications where submillisecond response times are typically required for safety-critical operations, and the average bandwidth of the testbeds shows that testbeds reliance on software defined network (SDN) in a virtualised environment can provide a near true to theoretical speed bandwidth (using gigabit Ethernet).
Average packet loss was measured to see whether there are packet drops in the network during normal load and during stress testing. From both measurements, no packet loss was detected, which shows that the platform has enough resources to accommodate the scenarios and that the network devices were configured correctly.

\begin{table}[htbp]
\caption{SIMPLE-ICS OT Protocols Performance Over One Hour}
\label{tab_ics_performance}
\centering
\small
\renewcommand{\arraystretch}{1.1}
\begin{tabularx}{\columnwidth}{l *{3}{>{\centering\arraybackslash}X}}
\toprule
\textbf{Metrics} & \textbf{Result}  \\
\midrule
Avg. number of Modbus Packets  &  1,389,194\\
Avg. number of MQTT Packets         & 39,274 \\
Avg. number of OPC UA Packets       & 33,045  \\

\bottomrule
\end{tabularx}
\end{table}

The use of testbed resources was evaluated across multiple dimensions to assess efficiency and scalability. The host system specifications included two servers interconnected within the DIATeam cyber-range virtual environment, each server housing an Intel Xeon Gold Processor with a total of 48 cores and 4 TB of RAM and NVMe storage, representing a realistic industrial virtualisation platform.

The CPU allocation for the testbed was 51 virtual machines using a total of 143 virtual processors. And CPU usage during normal operations was averaged 25\%, with peak usage reaching 37.9\% during the simulation of the APT scenario. Memory consumption remained stable at 146 GB for node 1( 6.5\% of total capacity), providing ample headroom for additional virtual machines or simulation complexity.



\subsection{Comparison with Similar Testbeds}

Table \ref{tab_testbed_comparison} compares SIMPLE ICS with contemporary testbeds across five dimensions: architecture (IT/OT/IIoT coverage), ICS fidelity (physical, hybrid, or digital twin), APT emulation support, key characteristics (flexibility, reproducibility, cost-effectiveness, isolation, openness \cite{green_ics_2020}), and data logging capabilities (system, network, operational logs).

\begin{table*}[ht]
\caption{SIMPLE ICS Comparison to other works identified in the literature}
\label{tab_testbed_comparison}
\centering
\renewcommand{\arraystretch}{1.2}

\resizebox{\textwidth}{!}{%

\begin{tabularx}{\textwidth}{p{3.5cm} p{2cm} p{0.25cm} p{0.25cm} p{0.3cm} p{2cm} p{0.25cm} p{0.25cm} p{0.25cm} p{0.25cm} p{0.25cm} p{0.25cm} p{0.3cm} p{0.3cm} p{0.3cm} }
\toprule
\multirow{2}{*}{\textbf{Testbed}} & \multirow{2}{*}{\textbf{ICS Fidelity}} & \multicolumn{3}{c}{\textbf{Architecture}} &\multirow{2}{*}{\textbf{APT}}  &\multicolumn{5}{c}{\textbf{Characteristics} } & & \multicolumn{3}{c}{Data Logs}\\
\cline{3-5} \cline{7-11} \cline{13-15}
&&\textbf{IT} &\textbf{OT} & \textbf{IIoT} & & \textbf{F} & \textbf{R} & \textbf{CE} &\textbf{I} &\textbf{O} & &Sys &Net &Op \\
\midrule
    BrownIIoTbed, 2021 \cite{al-hawawreh_developing_2021} & 
    Hybrid 
    & \ding{119} 
    & \ding{109}
    & \ding{108}
    & -
    & \ding{51}
    & \ding{51}
    & \ding{51}
    & \ding{51}
    & \ding{51}
    &
    & \ding{51}
    & \ding{51}
    & \ding{51}\\
\midrule
    WAMPAC, 2021 \cite{ravikumar_cps_2021} 
    & Hybrid 
    & \ding{109} 
    & \ding{108}
    & \ding{109}
    & -
    & \ding{53}
    & \ding{51}
    & \ding{53}
    & \ding{51}
    & \ding{53}
    &
    & \ding{53}
    & \ding{51}
    &\ding{51}\\
\midrule
    ICS Protable Cyber Kit, 2021 \cite{mubarak_industrial_2021} 
    & Scaled Down Physical  
    & \ding{109} 
    & \ding{108}
    & \ding{109}
    & -
    & \ding{53}
    & \ding{51}
    & \ding{53}
    & \ding{53}
    & \ding{53}
    &
    &\ding{53}
    & \ding{51}
    & \ding{51}\\
\midrule
    Karch et al., 2022 \cite{karch_crosstest_2022} 
    & Physical
    & \ding{119} 
    & \ding{108}
    & \ding{109}
    & -
    & \ding{53}
    & \ding{53}
    & \ding{53}
    & \ding{51}
    & \ding{53}
    &
    &\ding{53}
    & \ding{51}
    & \ding{53}\\
\midrule
    Simola et al., 2023 \cite{simola_developing_2023} 
    & Hybrid 
    & \ding{109} 
    & \ding{108}
    & \ding{119}
    & -
    & \ding{53}
    & \ding{51}
    & \ding{53}
    & \ding{51}
    & \ding{53}
    &
    &\ding{53}
    & \ding{51}
    & \ding{53}\\
\midrule
    Kumar and Thing, 2023 \cite{kumar_raptor_2023} 
    & Hybrid
    & \ding{119} 
    & \ding{109}
    & \ding{108}
    & Campaign
    & \ding{51}
    & \ding{51}
    & \ding{51}
    & \ding{51}
    & \ding{53}
    &
    &\ding{51}
    & \ding{51}
    & \ding{53}
    \\
\midrule
    Mikkelsplass \& Jorgensen, 2023 \cite{mikkelsplass_cyber_2023} 
    & Physical testbed 
    & \ding{109} 
    & \ding{109}
    & \ding{108}
    & -
    & \ding{53}
    & \ding{51}
    & \ding{51}
    & \ding{53}
    & \ding{53}
    &
    & \ding{53}
    & \ding{51}
    & \ding{53}
    \\
\midrule
    Ghiasvand et al, 2024 \cite{ghiasvand_cicapt-iiot_2024} 
    & Hybrid 
    & \ding{109} 
    & \ding{109}
    & \ding{108}
    & Stages
    & \ding{53}
    & \ding{51}
    & \ding{51}
    & \ding{53}
    & \ding{53}
    &
    & \ding{51}
    & \ding{51}
    & \ding{53}
    \\
\midrule 
    Lo et al., 2024 \cite{lo_digital_2024} 
    & Digital Twin 
    & \ding{109} 
    & \ding{108}
    & \ding{109}
    & -
    & \ding{53}
    & \ding{51}
    & \ding{51}
    & \ding{53}
    & \ding{53}
    & 
    & \ding{53}
    & \ding{53}
    & \ding{51}
    \\
\midrule
    SIMPLE ICS 
    & Digital Twin  
    & \ding{108} 
    & \ding{108}
    & \ding{108}
    & Campaigns
    & \ding{51}
    & \ding{51}
    & \ding{51}
    & \ding{51}
    & \ding{51}
    &
    & \ding{51}
    & \ding{51}
    & \ding{51}
    \\ 
\bottomrule
\end{tabularx}
}

\begin{tabularx}{\textwidth}{c c c c c}
    F: Flexibility & R: Reproduce & CE:Cost Effectiveness & I: Isolation & O: Openness\\
    \midrule
    \ding{108}: Implemented & \ding{119}: Partially Implemented & \ding{109}: Not Implemented &\ding{51}: Supported & \ding{53}: Not Supported 
    \\
    \bottomrule
\end{tabularx}
\end{table*}

\textbf{ICS Fidelity.} Physical testbeds \cite{karch_crosstest_2022,mubarak_industrial_2021,mikkelsplass_cyber_2023} offer high realism but are costly, difficult to reproduce, and require specialized environments. Hybrid testbeds \cite{al-hawawreh_developing_2021,simola_developing_2023} balance realism with reproducibility, though specialized components like RTDS simulators \cite{ravikumar_cps_2021} remain expensive. Digital twins \cite{lo_digital_2024,lazaridis_securing_2023} now provide sufficient fidelity for cybersecurity research while maximizing reproducibility and cost-effectiveness, motivating SIMPLE ICS's design choice.

\textbf{Architecture.} Most testbeds focus either on IIoT \cite{al-hawawreh_developing_2021,kumar_apt_2022,mikkelsplass_cyber_2023,ghiasvand_cicapt-iiot_2024} or traditional OT \cite{simola_developing_2023,ravikumar_cps_2021,mubarak_industrial_2021} architectures. Critically, despite documented APT attacks targeting traditional OT networks with IT infrastructure exploitation, no reviewed testbed fully recreates both IT systems (mail servers, Active Directory) and OT networks. SIMPLE ICS addresses this gap by supporting all three architectures.

\textbf{APT Emulation.} Only three testbeds explicitly support APT attacks.~\cite{kumar_apt_2022} and~\cite{karch_crosstest_2022} simulate APT scenarios but provide limited attack representations and no extensible framework. ~\cite{ghiasvand_cicapt-iiot_2024} focuses on individual attack stages rather than complete campaigns. Other testbeds use discrete, unrelated attacks that lack the sequential relationships characteristic of real APT campaigns. SIMPLE ICS provides a methodology and platform for emulating multi-stage APT campaigns with realistic attack chains from IT to OT/IIoT.

\textbf{Characteristics.} Physical testbeds lack flexibility and openness while incurring high costs. Hybrid approaches like~\cite{al-hawawreh_developing_2021} demonstrate good reproducibility and have enabled derivative works~\cite{kumar_apt_2022,ghiasvand_cicapt-iiot_2024}. Digital twins offer superior flexibility, reproducibility, and cost-effectiveness. Isolation is adequately addressed by most testbeds through virtualization or air-gapping, though \cite{mubarak_industrial_2021} and \cite{ghiasvand_cicapt-iiot_2024} lack explicit isolation documentation.

\textbf{Data Logging.} Network logs are nearly universal, but system and operational logs vary significantly. Only \cite{al-hawawreh_developing_2021} and SIMPLE ICS capture all three log types. System logs are essential for host-based detection but absent in most physical testbeds. Operational logs, critical for detecting attacks manifesting as physical anomalies, are surprisingly omitted by several IIoT-focused testbeds. SIMPLE ICS's comprehensive multi-source logging enables research on detection approaches that leverage correlated data across network, host, and operational domains.

\section{Limitations and Future Directions}
\label{sec_limitations}

Although SIMPLE ICS aims to drive advances in industrial cybersecurity testbed capabilities, several limitations remain that present opportunities for future enhancement.
\begin{itemize}
    \item Currently, the testbed's device simulation relies primarily on software-based industrial components, which, while functionally accurate, may not capture all the timing characteristics and physical constraints of actual industrial hardware. Future work will explore hybrid approaches that integrate hardware-in-the-loop components to further increase realism.
    \item The testbed's current focus on traditional industrial protocols, while comprehensive for most scenarios, could be expanded to include emerging IIoT protocols and edge computing architectures that are increasingly prevalent in modern industrial environments. In addition, the integration of artificial intelligence and machine learning components for both attack simulation and defence mechanisms represents a promising research direction.
    \item To further enhance the framework's research capabilities, our immediate development roadmap includes the implementation of two additional APT attack scenarios that will significantly expand the range of Tactics, Techniques, and Procedures (TTPs) available for simulation. These new scenarios will complement our existing Sandworm, BlackEnergy, and FrostyGoop implementations by targeting different attack vectors and industrial sectors, providing researchers with a more comprehensive representation of the APT threat landscape. Concurrently, we will generate a comprehensive APT attack dataset from these expanded simulations, capturing network traffic, system logs, and behavioural patterns across all implemented attack scenarios. This data set will serve as a valuable resource for the research community, enabling the development and validation of machine learning-based detection systems, behavioural analysis algorithms, and threat intelligence frameworks specifically tailored to industrial environments.
\end{itemize}
\section{Conclusions}
\label{sec_conclusions}
This paper has presented SIMPLE-ICS (Simulated Industrial Multi-tier Platform for Laboratory Emulation of Industrial Control Systems), a testbed framework designed to be easily reproduced by cybersecurity researchers. Through systematic design, implementation, and validation, we have demonstrated that SIMPLE-ICS provides a realistic and extensible platform for simulating Advanced Persistent Threat attacks against modern industrial enterprise networks.
The key contributions of this work are as follows. Firstly, we introduced an adaptation of the V-Model design methodology for testbed development. Secondly , we presented a novel multitier framework architecture that seamlessly integrates IT, OT, and IIoT components within a virtualized environment, allowing researchers to study cyber-physical security interactions across the entire industrial enterprise stack. The modular design, implemented on the DIATeam cyber range infrastructure with standardised industrial protocols (Modbus, OPC UA, and MQTT), ensures both realism and extensibility for diverse industrial scenarios.
Third, we systematically recreate real-world APT attack campaigns that begin by developing a generic attack scenario distilled from prominent APT attack incidents. We then modelled the attack by mapping the scenario to the MITRE ATT\&CK framework, and implemented the attack with the Caldera platform augmented by several manual attacks such as delivering the initial payload (initial access). By accurately modelling the Tactics, Techniques and Procedures (TTPs) of prominent threat actors including Sandworm, BlackEnergy, and FrostyGoop, SIMPLE ICS enables researchers to conduct realistic adversarial simulations that closely mirror actual industrial cyber threats. This capability addresses a significant limitation in existing testbeds that often rely on simplified or theoretical attack scenarios.
And last, we established a comprehensive validation framework that demonstrates our simulated industrial environments through network behaviour analysis, protocol communication verification, and attack scenario validation.

Our evaluation results show that SIMPLE-ICS achieves high realism in network traffic patterns and industrial process behaviours while maintaining the flexibility needed for diverse research applications.

SIMPLE-ICS is envisioned to become a collaborative ecosystem in which researchers and practitioners contribute new attack scenarios, industrial configurations, and validation methodologies. 
In conclusion, SIMPLE ICS represents a step forward in the industrial cybersecurity research infrastructure, providing an extensible and realistic framework to advance our understanding of cyber threats to critical infrastructure. Through continued community collaboration and development, we anticipate that SIMPLE ICS will play a role in strengthening the cybersecurity of industrial systems worldwide, ultimately contributing to the protection of critical infrastructure that underpins modern society.



\bibliographystyle{spphys}
\bibliography{bib/references}

\end{document}